\documentclass[aps,prl,preprint,12pt]{revtex4}
\usepackage{graphicx}
\begin{document}
\title{Anisotropic magnetoresistance of charge-density wave in $o$-TaS$_3$}
\author{Katsuhiko Inagaki, Toru Matsuura,$^{1,2}$ Masakatsu Tsubota,$^{3}$ Shinya Uji,$^4$ Tatsuya Honma, 
and Satoshi Tanda$^{1,2}$}
\affiliation{Department of Physics, Asahikawa Medical University, 2-Jo 1-Chome, Midorigaoka-Higashi, Asahikawa, 078-8510 Japan}
\affiliation{$^1$Division of Applied Physics, Hokkaido University, Kita 13 Nishi 8, Kita-ku, Sapporo 060-8628, Japan}
\affiliation{$^2$Center of Education and Research for Topological Science and Technology,
Hokkaido University, Kita 13 Nishi 8, Kita-ku, Sapporo 060-8628, Japan}
\affiliation{$^3$Division of Quantum Science and Engineering, Hokkaido University, Kita 13 Nishi 8, Kita-ku, Sapporo 060-8628, Japan}
\affiliation{$^4$National Institute for Materials Science,
1-2-1 Sengen, Tsukuba 305-0047, Japan}

\begin{abstract}
We report the magnetoresistance of a charge-density wave (CDW) in $o$-TaS$_3$ whiskers at 4.2 K under a magnetic field up to 5.2 T.
An anisotropic negative magnetoresistance is found in the nonlinear regime of current-voltage characteristics.
The angle dependence of the magnetoresistance, studied by rotating the magnetic field upon the $c$-axis, exhibited a two-fold symmetry.
The magnetoresistance amplitude exhibits maxima when the field is parallel to the $a$-axis, whereas it vanishes to the $b$-axis.
The observed anisotropy may come from difference in interchain coupling of adjacent CDWs along the $a$- and $b$-axes.  
Comparison of the anisotropy to the scanning tunneling microscope image of CDWs allows us to provide a simple picture to explain the magnetoresistance in terms of delocalization of quantum interference of CDWs extending over the $b$-$c$ plane.
\end{abstract}
\maketitle

Dynamics of charge density waves (CDWs) has been of interest for decades \cite{Gruner}.
In particular, effect of a magnetic field to CDW dynamics remains unsolved.
Phase of a CDW determines the initial position of the density wave,
and collective motion of the CDW is described in terms of variation of the phase. 
Naively, no magnetic responses would be expected since CDWs consist of hole and electron pairs, whose charge is neutral. 
However, Aharonov-Bohm (AB) oscillations of CDWs have been reported in two systems, firstly in 
the ion-beam-radiated NbSe$_3$ \cite{Latyshev1997}, then in the ring crystals of TaS$_3$ \cite{Tsubota2012}.
The observed oscillation period was $h/2e$ for the both cases, suggesting that quantum interference might occur for the phase of a CDW with the charge of $2e$, which coupled with a vector potential. 
Theoretical interpretation of the AB oscillations in CDWs has not been established.
Direct coupling between the CDW phase and a vector potential was firstly proposed \cite{Bogachek1990}, then CDW solitons were studied as carriers affected by a magnetic field \cite{Tsubota2012,Miller2013}.
Therefore, it is important to focus on magnetoresistance of CDWs in a trivial topology such as a simple whisker in order to reveal the mechanism of such the quantum interference. To study the magnetoresistance of CDWs, systems with imperfect nesting should be avoided because it
might hinder possible effects to CDW dynamics. 
For example, NbSe$_3$, which is known as an imperfect nesting CDW system, exhibits a large positive magnetoresistance at low temperatures \cite{Tritt1991}, resulting from uncondensed electrons remained on the Fermi surface even below the Peierls temperature.

In this article, we report the magnetoresistance of a CDW in $o$-TaS$_3$ whiskers.
The whole Fermi surface of $o$-TaS$_3$ disappears
below its Peierls temperature, hence at low temperatures electric conduction is only due to the CDW.
We measured the resistance in the nonlinear regime
of $o$-TaS$_3$  under magnetic field up to 5.2 T at 4.2 K, which is low enough to prevent from thermally exciting quasiparticles. 
The magnetoresistance was negative in sign.
The angle dependence of the magnetoresistance was studied by rotating the magnetic field upon the $c$-axis, namely,  the chain axis of the
crystal.
A two-fold symmetry was found in the magnetoresistance.
The magnetoresistance amplitude exhibits maxima when the field is parallel to the $a$-axis, whereas it vanishes to the $b$-axis.
The observed anisotropy may come from difference in interchain coupling of adjacent CDWs along the $a$- and $b$-axes.
Comparison of the anisotropy to the scanning tunneling microscope (STM) image of CDWs allows us to provide a simple picture to explain the magnetoresistance in terms of delocalization of quantum interference of CDWs extending over the $b$-$c$ plane.
Our observation is an unexpected and new phenomenon, which will provide
an important key to understand the AB oscillations of the CDWs.

Single crystals of $o$-TaS$_3$ were grown using a standard chemical vapor
transportation method.
A pure tantalum sheet and sulfur powder were placed in a
quartz tube. The quartz tube was evacuated to $1 \times 10^{-6}$ Torr
and heated in a furnace at 530 ${}^\circ$C for two weeks. 
The grown crystals were ribbon-like whiskers. The chain direction of $o$-TaS$_3$ is 
along to the $c$-axis, and the flat surface of the ribbon is reported to be 
a $b$-$c$ plane, perpendicular to the $a$-axis
 \cite{Tsutsumi1978,Sugai1984,Gammie1989}.
The crystal orientation of each sample was determined and confirmed to be consistent with the previous reports by the
electron back scattering diffraction (EBSD) technique (OIM TSL).
The electrodes were made using 50-$\mu$m-diameter silver wires glued with
silver paint. Gold thin film was deposited on the crystal before
the silver wires were attached to reduce the contact resistance
to 1 $\Omega$ at room temperature.

The resistance of the sample was measured with a standard four-probe technique.
As described in a previous study \cite{Inagaki2010}, a high-impedance digital voltmeter (Keithley 6512; $Z_\mathrm{in}> 200$ T$\Omega$) was employed. 
All measurements were performed with constant currents
generated by a current source (Keithley 220). 
A magnetic field was applied with a couple of superconducting coils.
The sample holder was rotated along an axis perpendicular
to the magnetic line. In this experiment
the axis of rotation was aligned with the chain axis of the sample.
The sample was glued to the sample holder with the ribbon surface
facing the holder. 
Since Joule heat induced by eddy current might have caused the temperature to increase when the holder was rotated, each measurement was performed after the temperature was stabilized in less than 4 mK rise.

\begin{figure}
\includegraphics[width=0.9\textwidth]{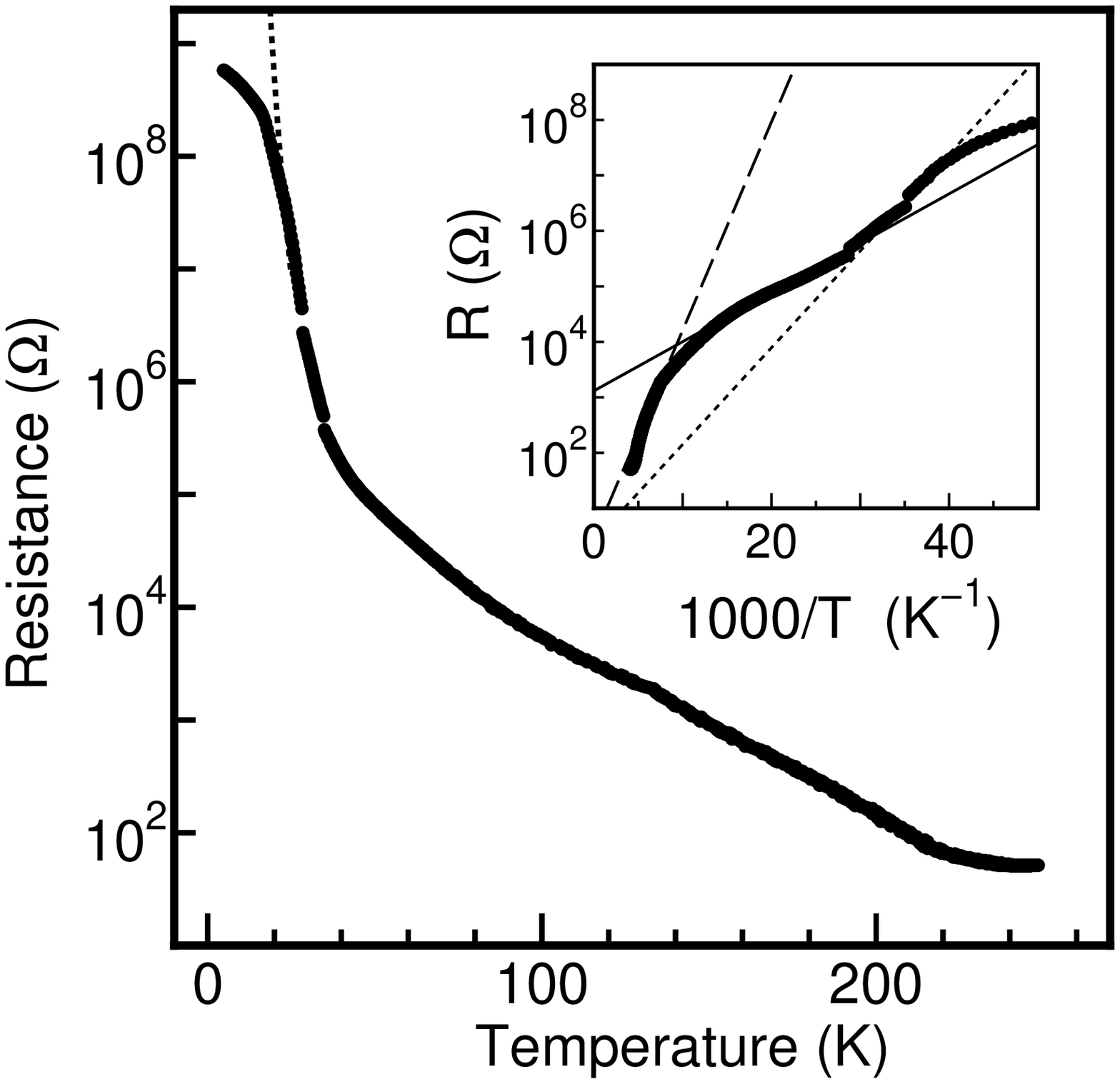}

\caption{Temperature dependence of $o$-TaS$_3$. The discontinuity is
caused by a change in the measurement current. Inset shows three Arrhenius fits for high, intermediate, and low temperature ranges, denoted by broken, sold, and dotted lines, respectively. The low temperature fit is also shown in the main panel by the dotted line, by which the extrapolation of the ohmic resistance at 4.2 K appears to exceed $10^9\ \Omega$.}
\label{fig_rt}
\end{figure}

Figure \ref{fig_rt} shows the typical temperature dependence of resistance ($R$-$T$).
The sample cross-section is $15 \times 0.5$ $\mu\mathrm{m}^2$.
The room temperature resistivity
of the sample is $2.8 \times 10^{-6}\ \Omega \mathrm{m}$, which is consistent
with previous reports ($\sim 3 \times 10^{-6}\ \Omega \mathrm{m}$) \cite{Takoshima1980,Staresinic2002}.
By lowering the temperature, the system undergoes a Peierls transition
at 220 K, below which the electrons at the Fermi surface condense into
a charge density wave state. In the 100 to 200 K temperature range,
the resistance obeys an Arrhenius law with an activation energy of 860 K (broken line in the inset of Fig. \ref{fig_rt}).
Discrepancy from the Arrhenius law is found below 100 K. 
In the 40 to 100 K temperature range, a smaller activation
energy ($\sim 200$ K; solid line) is applicable, and it becomes higher ($\sim 400$ K; dotted line) at temperatures below 30 K.
Such behavior is reproducible and also consistent with previous reports \cite{Zhilinskii1983,Staresinic2002}.

\begin{figure}
\includegraphics[width=0.9\textwidth]{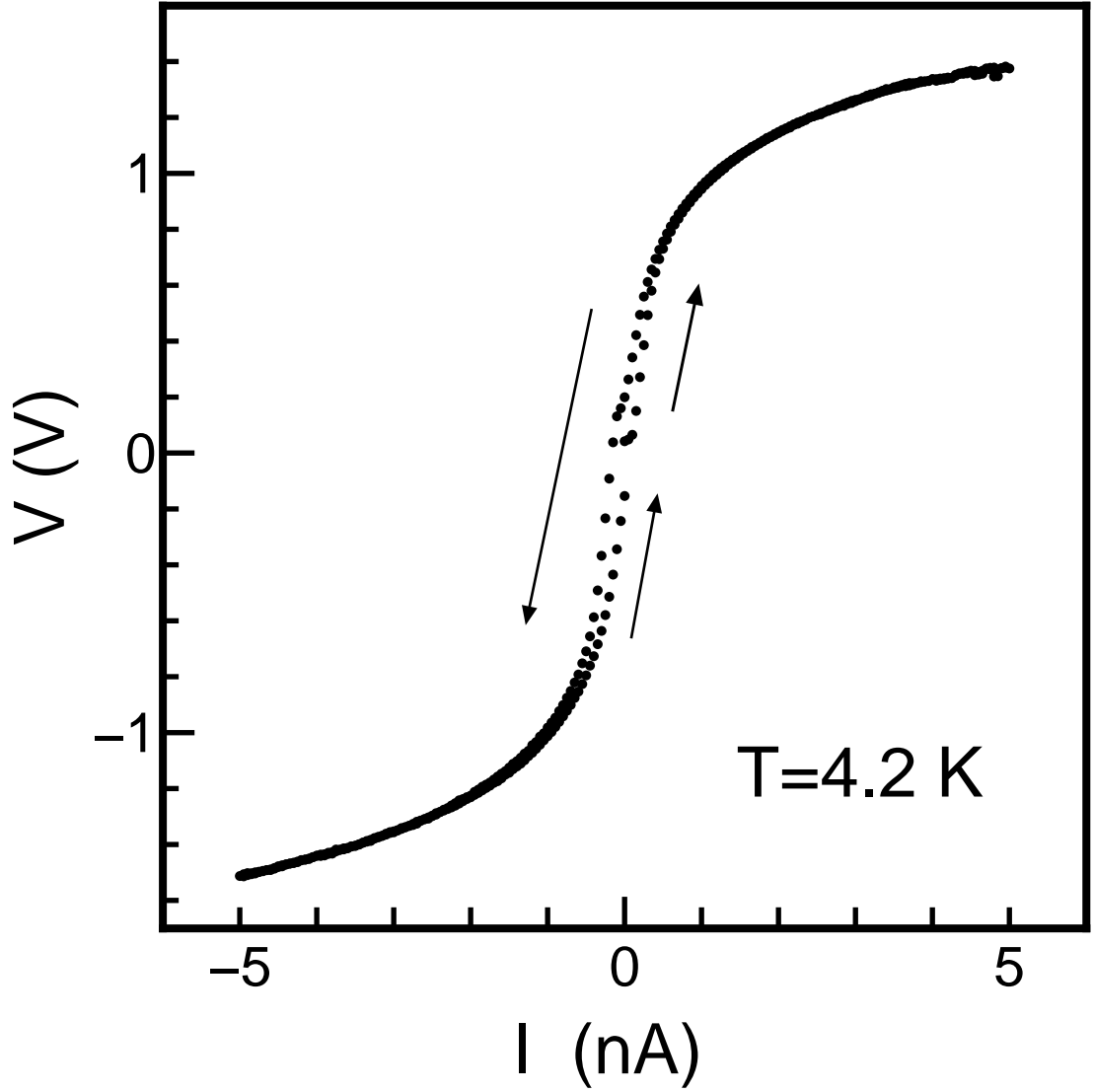}
\caption{Current-voltage ($I$-$V$) characteristics of the $o$-TaS$_3$ sample
observed at 4.2 K. The $I$-$V$ curve is significantly nonlinear.
A slight hysteresis is found near $I$=0 as shown by the arrows.}
\label{fig_iv}
\end{figure}

Figure \ref{fig_iv} shows the current-voltage ($I$-$V$) characteristics of the sample at 4.2 K with flowing current $I$=2 nA.
At this temperature, the ohmic resistance exceeds $10^9$ $\Omega$, as deduced from
an extrapolation of the $R$-$T$ curve.
However, the slope 
of the $I$-$V$ curve corresponds to $1.3 \times 10^9$ $\Omega$, which comes
from a tiny current accompanied by relaxation.
No nonlinear conduction threshold was observed in the
$I$-$V$ characteristics at this temperature where there were almost no thermally activated quasiparticles.
A slight hysteresis was also observed in the $I$-$V$ curve in the neighborhood
of $I$=0. This phenomenon can be interpreted as a rearrangement of 
the CDW dislocations, which hold electric charges, as reported for blue bronze \cite{Tessema1985,Zybtsev2010}, and  TaS$_3$ \cite{Borodin1988}. 
In a higher current range $|I| > 1 \times 10^{-9}$ A, the hysteresis became
insignificant. The following experiments were
performed in this current range.

\begin{figure}
\includegraphics[width=0.9\textwidth]{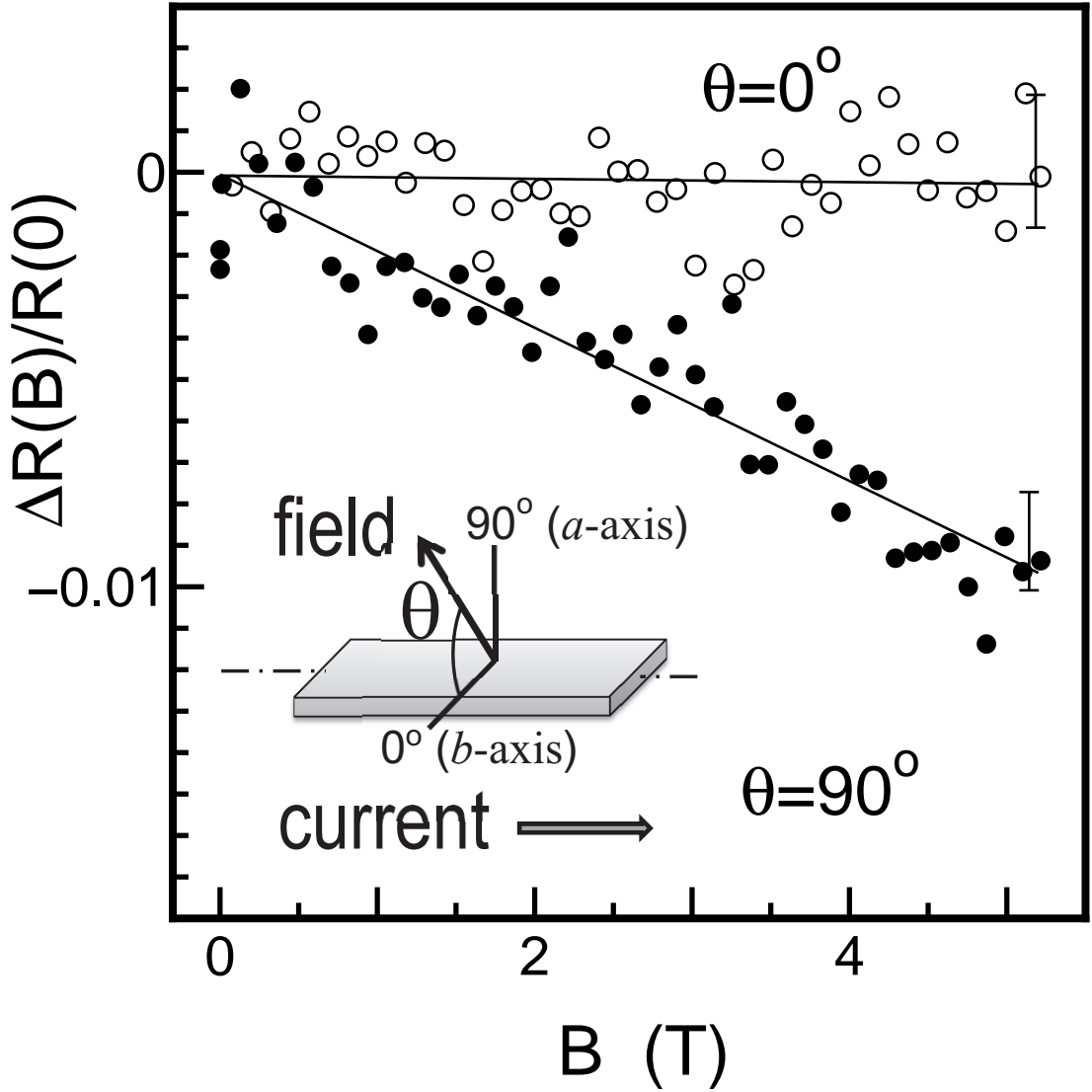}
\caption{Magnetoresistance of the $o$-TaS$_3$ sample observed at 4.2 K.
$\theta=0$ (open circles) means the magnetic field is parallel to the current flow,
and at $\theta=90^\circ$ (solid circles) the field is perpendicular to the ribbon face, as shown in the inset. The current is 2 nA, which stays in the nonlinear regime of the $I$-$V$ characteristics. The error bar represents $\pm 2\sigma$, where $\sigma$ is the standard deviation of the data. Solid lines are the guide to the eyes.}
\label{fig_mag}
\end{figure}

Figure \ref{fig_mag} shows the magnetoresistance of the sample observed at 4.2 K.
The magnetic field is applied perpendicular to the current flow ($c$-axis). 
$\theta=0^\circ$ means that the field is directed along the $a$-axis,
and at $\theta=90^\circ$ the field is along the $b$-axis, as shown in the inset.
This result reveals negative magnetoresistance with anisotropy,
and is reproduced in several samples. We also confirmed that the observed
magnetoresistance is independent of the current direction.
Moreover, the ratio $V(B)/V(0)$ was constant  over the nonlinear regime of the $I$-$V$ characteristics in the field of $B=5.2$ T.
As shown by the error bars in Fig. \ref{fig_mag}, which correspond to $\pm 2\sigma$ where $\sigma$ is a standard deviation of raw data, the data seem to be noisy, probably because of influence of the CDW's collective motion, e. g. narrow band noise.

Magnetoresistance of quasi one-dimensional conductors has been intensively studied \cite{Blundell1996}. 
A magnetic field changes electron motion on the Fermi surface. This provides an increase of resistivity, namely, \textit{positive magnetoresistance} with anisotropic angle dependence according to the shape and topology of the Fermi surface.
On the contrary, the sign of the magnetoresistance of $o$-TaS$_3$ was \textit{negative}  (Fig. \ref{fig_mag}). 
Moreover, 
no normal carrier in $o$-TaS$_3$ is left at the Fermi surface in the CDW state,
and thermally activated quasiparticles are negligible at 4.2 K. Hence, our observation should not be understood in terms of the conventional magnetoresistance.
A comparison with the magnetoresistance of the NbSe$_3$ case \cite{Tritt1991} is also noteworthy. NbSe$_3$ has two CDW transitions at $T_1= 145$ K and $T_2 = 59$ K. Even below $T_2$ there remain normal carriers on the Fermi surface.
A large positive magnetoresistance was observed below $T_2$, accompanied with 
increase of the number of CDW carriers. Magnetic response to the dynamics of a CDW was hindered in the previous experiment. 
Our result also excludes the possibility of a spin-related
phenomenon being a major contributor to the observed magnetoresistance, as with the negative magnetoresistance of TaS$_2$ \cite{Kobayashi1979}, which is essentially isotropic. 

\begin{figure}
\includegraphics[width=0.9\textwidth]{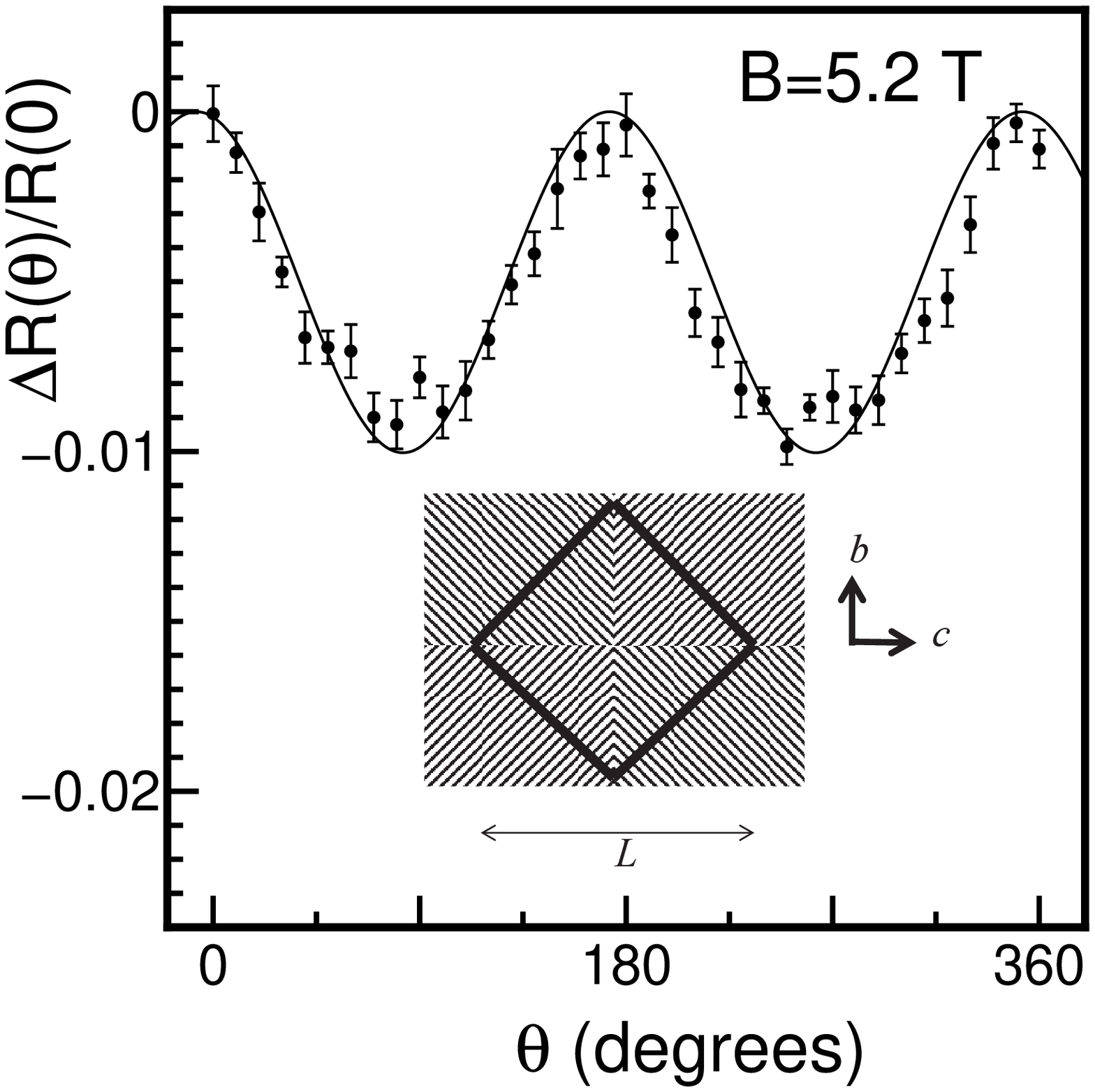}
\caption{Angle dependence of the magnetoresistance in the field $B=5.2$ T observed at 4.2 K. The error bars represent $\pm 2\sigma$ as in Fig. \ref{fig_mag}. A two-fold symmetry is 
clearly exhibited. The solid line shows a result of a least-square fit to the formula $\Delta R = - A\sin ^2(\theta-\theta_0)$, where the offset angle $\theta_0=-7^\circ$. This angle coincides with the direction of $b$-axis of the sample determined with the EBSD technique. The inset shows a schematic image of a CDW loop formed in domain structure. The thin lines represent wavefront of CDWs, and the bold line shows a loop, surrounded an area of $L^2\sin\alpha$, where $L$ is the length of the loop, and $\alpha$ is the angle between CDW and the $b$-axis. The angle $\alpha$ is exaggerated in the inset. }
\label{fig_angle}
\end{figure}

Figure \ref{fig_angle} shows the angle dependence of the magnetoresistance,
which reveals the two-fold symmetry of the magnetoresistance.
As the first approximation for two-fold symmetry,
we tried to fit the angle dependence of the observed magnetoresistance with the formula:
$\Delta R = - A \sin ^2 (\theta-\theta_0)$, where $A$ is the amplitude of magnetoresistance and $\theta_0$ is the offset
angle. 
The parameters were determined by nonlinear least square fitting to be 
$A= 5.78 \times 10^6$~$\Omega$ and $\theta=-7^\circ$.
The solid line in Fig. \ref{fig_angle} shows the result of the fit.
The residual error of the fit is roughly the same as the distribution of the observed data, which is shown with the error bars of 2$\sigma$ in Fig. \ref{fig_angle}.
The magnitude of the magnetoresistance exhibits maxima when the field is applied along the $a$-axis, whereas it vanishes when the field is along the $b$-axis.
The offset angle $\theta_0=-7^\circ$ represents a slight misalignment of the crystal axes to the magnetic field direction. This angle coincides with the direction of the $b$-axis of the sample determined with the EBSD technique.

It is necessary to look carefully at difference between the $a$- and $b$- axes. 
The STM image \cite{Gammie1989} demonstrates that maxima of CDW intensity on the $b$-$c$ plane are canted and form an angle of 86$^\circ$ with the chain axis. 
This angle corresponds to $\mathrm{arctan}(8b_0/4c_0)\sim 84^\circ$, where $b_0$ and $c_0$ are the lattice constants of $o$-TaS$_3$.
This implies the wavefunction of CDW extends over the $b$-$c$ plane,
and it is not firmly bound to a particular chain.
The STM image is consistent with the CDW vector of $o$-TaS$_3$, $\mathbf{q}_\mathrm{CDW}=(0.5, 0.125, 0.25)$, formerly determined with diffraction studies \cite{Tsutsumi1978,Roucau1983}.
CDWs of adjacent chains are out of phases along the $a$-axis. This results from energetically favorable configuration among independent chains coupled with the Coulomb interaction \cite{Gruner}.
Hence the response to the magnetic field may \textit{not} be symmetric for the rotation upon the $c$-axis. 

The canting of CDW wavevector provides a simple picture as follows. There are two possibilities for the CDW to settle on the pristine lattice by canting left or right. 
Once either left or right is chosen, a certain area of CDWs are aligned to form a domain. If CDWs consist of such the domains,
a closed loop can exist as shown in the inset of Fig. \ref{fig_angle}.
 The area surrounded by the loop is represented as $S=L^2 \sin \alpha$, where $L$ is the coherent length along the $c$-axis, and $\alpha$ is the angle between CDW wavefront and the $b$-axis.
In fact, the size of domains is distributed and such the loops may be formed randomly in the sample.
Self-interference of a wavefunction in a random media, as known as the Anderson localization \cite{Altshuler1980}, can explain the magnetic response. 
A magnetic field destroys the interference of waves which go around the close loop. This phenomenon is known as delocalization and provides a negative magnetoresistance \cite{Sasaki1965,Tanda1991}. 
A relative shift of the electron phase $\Delta\psi$ is represented as $\Delta\psi=\frac{e}{h}\int{\bf A} d{\bf l}$,
where $h$ is Planck's constant, $e$ is the charge quantum of the carrier, and $\bf A$
is the vector potential. 
In a two-dimensional system, electrons can only move on a conduction plane.
With a closed path, which surrounds an area $S$,
the phase shift is proportional to the flux $\Phi=BS$
where ${\bf B} = \mathrm{rot} {\bf A}$ is an applied field. 
Since Anderson localization results from the superposition of possible closed paths of electrons, this leads to the anisotropy of the magnetoresistance when the applied field
is inclined by an angle $\theta$ to the conducting plane, and
the effective area becomes $S \sin \theta$. Without spontaneous magnetization \cite{remark1}, magnetoresistance is
represented as an even function of the magnetic flux, e.g. $\Delta R \propto - \Phi^2$, hence magnetoresistance would have a component of $\sin ^2\theta$.
Therefore, the observed two-fold symmetry in magnetoresistance (Fig. \ref{fig_angle}) can be interpreted as  a natural consequence
of the delocalization picture.

The coherent length $L$ is estimated both by the STM image and by X-ray diffraction. The STM image (Fig. 3 in Ref. \cite{Gammie1989}) shows CDWs in $22.5 \times 22.5$ nm$^2$ area, where two domains of CDW can be distinguished. This image suggests that the size of each domain is much larger than $2\times 10^{-8}$ m.
A synchrotron X-ray study \cite{Inagaki2008} exhibits the coexistence of incommensurate and commensurate CDWs in $o$-TaS$_3$. Two satellite spots are separated by two pixels apart at the detector, and each spot was as narrow as one-pixel width (Fig. 1 in Ref \cite{Inagaki2008}). This gives an estimation of the correlation length longer than $3 \times 10^{-7}$ m in $c^*$-direction.
If $L\sim 3\times 10^{-7}$ m is assumed, the corresponding area of the CDW loop becomes $1 \times 10^{-14}$ m$^2$, which gives the field of 0.2 T for a flux quantum $h/2e$.
This field can be interpreted as the minimum field for the negative magnetoresistance to occur. 
The observed magnetoresistance ranges above the field of 0.2 T, as shown in Fig. \ref{fig_mag}, hence it is consistent quantitatively with the delocalization picture.

What kinds of the carrier are consistent with the delocalization picture? 
Quantum interference can occur for any kind of carriers as far as their phase is not lost.
We have already ruled out magnetoresistance of thermally activated quasiparticles as a possible candidate for our observation. 
If the collective motion of CDW plays a major role for the observed negative magnetoresistance, such the quantum interference would increase pinning rate of the CDW. 
In the case of soliton transport, a soliton travels along such the closed path may be inactive for a carrier.
The negative magnetoresistance with the two-fold symmetry is consistent for the both cases.

Finally, our observation and interpretation will ease conditions to observe the AB oscillations in CDWs. 
The previous studies were performed with the ion-beam-radiated NbSe$_3$ \cite{Latyshev1997} and the ring crystal of TaS$_3$ \cite{Tsubota2012}. 
Tsubota \textit{et al.} proposed that a CDW soliton might be confined and move along a single chain whose ends coalesced \cite{Tsubota2012}. 
This proposal was based on the growth mechanism of the ring crystals \cite{Tanda2002}.
However, the magnetoresistance is interpreted as quantum interference of the CDWs extending over the $b$-$c$ plane, and CDWs can maintain their quantum phase across the chains.
Therefore, our observation is not only consistent with the previous reports,
but also suggesting possibility of the AB oscillations in a closed CDW loop realized by other methods.

We are grateful to K. Yamaya, S. Takayanagi, and K. Ichimura for fruitful discussions, and to T. Ikeda and T. Kanno for experimental support.

\clearpage



\begin{thebibliography}{99}
\bibitem{Gruner}
G. Gr\"uner, \textit{Density waves in solids}, Addison-Wesley Longmans, Reading (1994).
\bibitem{Latyshev1997}
Yu. I. Latyshev, O. Laborde, P. Monceau, and S. K. Klaumunzer,
Phys. Rev. Lett. \textbf{78}, 919 (1997).
\bibitem{Tsubota2012}
M. Tsubota, K. Inagaki, T. Matsuura, and S. Tanda,
Europhys. Lett. \textbf{97}, 266404 (2012).
\bibitem{Bogachek1990}
E.N. Bogachek, I.V. Krive, I.O. Kulik, and A.S. Rozhavsky,
Phys. Rev. B \textbf{42}, 7614 (1990).
\bibitem{Miller2013}
J. H. Miller, Jr., A. I Wijesinghe, Z. Tang, and A. M. Guloy,
Phys. Rev. B \textbf{87}, 115127 (2013).
\bibitem{Tritt1991}
T. M. Tritt, A.C. Ehrlich, D. J. Gillespie, and G. X. Tessema,
Phys. Rev. B \textbf{43}, 7254 (1991).
\bibitem{Tsutsumi1978}
K. Tsutsumi, T. Sambongi, S. Kagoshima, T. Ishiguro,
J. Phys. Soc. Jpn \textbf{44}, 1735 (1978).
\bibitem{Sugai1984}
S. Sugai, Phys, Rev. B \textbf{29}, 953 (1984).
\bibitem{Gammie1989}
G. Gammie, J. S. Hubacek, S. L. Skala, R. T. Brockenbrough, J. R. Tucker, and J. W. Lyding,
Phys. Rev. B \textbf{40}, 11965 (1989). 
\bibitem{Inagaki2010}
K. Inagaki, M. Tsubota, and S. Tanda,
Phys. Rev. B \textbf{81}, 113101 (2010).
\bibitem{Staresinic2002}
D. Stare\v{s}ini\'c, K. Biljakovi\'c, W. Br\"utting, K. Hosseini, P. Monceau, H. Berger, and F. Levy, Phys. Rev. \textbf{B 65}, 165109 (2002).
\bibitem{Takoshima1980}
T. Takoshima, M. Ido, K. Tsutsumi, T. Sambongi, S. Honma, K. Yamaya, and Y. Abe,
Solid State Communications \textbf{35}, 911 (1980).
\bibitem{Zhilinskii1983}
S.K. Zhilinski\u\i, M.E. Itkis, I.Yu. Kal'nova, F.Ya. Nad', and V.B. Preobrazhenski\u\i, Sov. Phys. JETP \textbf{58}, 211 (1983).
\bibitem{Tessema1985}
G. X. Tessema, B. Alavi, and L. Mihaly,
Phys. Rev. B \textbf{31}, 6878 (1985).
\bibitem{Zybtsev2010}
S.G. Zybtsev, V.Ya. Pokrovskii, and S.V. Zaitsev-Zotov,
Nature Commun. \textbf{1}, 85 (2010).
\bibitem{Borodin1988}
D.V. Borodin, S. V. Za\u{\i}tsev-Zotov, and F. Ya. Nad',
Sov. Phys. JETP \textbf{66}, 793 (1988).
\bibitem{Blundell1996}
S.J. Blundell and J. Singleton,
Phys. Rev. B \textbf{53}, 5609 (1996).
\bibitem{Kobayashi1979}
N. Kobayashi and Y. Muto,
Solid State Commun. \textbf{30}, 337 (1979).
\bibitem{Roucau1983}
C. Roucau,
J. Phys. (Paris) \textbf{44}, 1725 (1983).
\bibitem{Altshuler1980}
B. L. Altshuler, D. Khmel'nitzkii, A. I. Larkin, and P. A. Lee,
Phys. Rev. B \textbf{22}, 5142 (1980).
\bibitem{Sasaki1965}
W. Sasaki, J. Phy. Soc. Jpn. \textbf{20}, 825 (1965). 
\bibitem{Tanda1991}
S. Tanda, M. Honma, and T. Nakayama,
Phys. Rev. B \textbf{43}, 8725 (1991).
\bibitem{remark1}
Our observation does not exclude the possibility of spontaneous magnetization.
If a spin density wave is locally accompanied with a CDW around a pinning center,
such a possibility should be considered. I. T\"utt\"o and A. Zawadowski,
Phys. Rev. B \textbf{32}, 2449 (1985).
\bibitem{Inagaki2008}
K. Inagaki, M. Tsubota, K. Higashiyama, K. Ichimura, S. Tanda, K. Yamamoto, N. Hanasaki, N. Ikeda, Y. Nogami, T. Ito, and H. Toyokawa,
J. Phys. Soc. Jpn. \textbf{77}, 093708 (2008).
\bibitem{Tanda2002}
S. Tanda, T. Tsuneta, Y. Okajima, K. Inagaki, K. Yamaya, and N. Hatakenaka,
Nature \textbf{417}, 397 (2002).
\end{thebibliography}
\end{document}